\documentclass[preprint,showpacs,preprintnumbers,amsmath,amssymb]{revtex4}

\usepackage{graphicx}
\usepackage{dcolumn}
\usepackage{bm}
\usepackage{booktabs}

\begin{document}

\title{\large Observation of two distinct pairs fluctuation lifetimes and\\
 supercurrents in the pseudogap regime of cuprate junctions
 }

\author{Gad Koren}
\email{gkoren@physics.technion.ac.il} \affiliation{Department of Physics,
Technion - Israel Institute of Technology, Haifa
32000, ISRAEL} \homepage{http://physics.technion.ac.il/~gkoren}

\author{Patrick A. Lee}
\affiliation{Department of Physics, Massachusetts Institute of Technology,
Cambridge, Massachusetts 02139, USA}

\date{\today}
\def\bfig {\begin{figure}[tbhp] \centering}
\def\efig {\end{figure}}

\normalsize \baselineskip=8mm  \vspace{15mm}


\begin{abstract}

The origin of the pseudogap in the high temperature superconductors is still unknown. Proposals for pairing in this regime range from the existence of preformed pairs which are believed to be precursors to superconductivity, to various competing orders such as charge and pair density waves. Here we report on pairs fluctuation supercurrents and inverse lifetimes in the pseudogap regime. These were measured on epitaxial \textit{c-axis} junctions of the cuprates, with a $PrBa_2Cu_3O_{7-\delta}$ barrier sandwiched in between two $YBa_2Cu_3O_{7-\delta}$ or doped $YBa_2Cu_3O_y$ electrodes, with or without magnetic fields parallel to the \textit{a-b} planes. All junctions had a $\rm T_c(high)\approx 85-90$ K and a $\rm T_c(low)\approx 50-55$ K electrodes, allowing us to study pairs fluctuation supercurrents and inverse life times in between these two temperatures. In junctions with a pseudogap electrode under zero field, an excesss current due to pair fluctuations was observed which persisted at temperatures above $\rm T_c(low)$, in the pseudogap regime, and up to about $\rm T_c(high)$. No such excess current was observed in junctions without a pseudo-gap in the electrode. The measured conductance spectra at temperatures above $\rm T_c(low)$ were fitted using a modified fluctuations model by Scalapino [Phys. Rev. Lett. \textbf{24}, 1052(1970)] of a junction with a serial resistance. We found that in the pseudo-gap regime, the conductance vs voltage consists of a narrow peak sitting on top of a very broad peak.  This yielded two distinct pairs fluctuation lifetimes in the pseudogap electrode which differ by an order of magnitude up to about $\rm T_c(high)$. Under in-plane fields, these two lifetime values remain separated in two distinct groups, which varied with increasing field moderately.  We also found that detection of Amperian pairing [Phys. Rev. X \textbf{4}, 031017 (2014)] in our cuprate junctions is not feasible, due to Josephson vortices penetration into the superconducting electrodes which drove the necessary field above the depairing field.\\

\end{abstract}

\maketitle

\section{Introduction}

More than 25 years after the discovery of the pseudogap in the cuprates \cite{Timusk-Statt,Norman,KeimerRev} its origin is still controversial. At the beginning, the pseudogap was referred to as a cross-over phenomena where many physical parameters have changed significantly at a temperature $T^*$ higher than $T_c$.  Later, experimental evidence was found that showed a real phase transition at $T^*$ \cite{Fauque,Shekhter}. An important question is whether the pseudogap is related to superconducting pair fluctuations. In some experiments diamagnetic response was observed up to 120 K \cite{LiLu} and inter-layer pair coherence up to 180 K \cite{Dubroka}, which suggest that strong pair fluctuations  possibly associated with the pseudo-gap, while other experiments indicate that pair fluctuations are limited to 20 K or so above $T_c$ \cite{Orenstein}, and maybe even Gaussian in nature \cite{Kokanovic}. A related question is whether a vortex liquid state exists much above the transport $H_{c2}$ which develops a strong dip near 1/8 doping \cite{Griss}. These authors argue that there is no vortex liquid state in the low temperature limit and therefore the true $H_{c2}$ is lower than or equal to the zero temperature transport $H_{c2}$, but other authors strongly disagree with this conclusion \cite{Ong}.  Another point of view is that the pseudogap is a competing or coexisting order with superconductivity \cite{Keimer,KeimerRev}. Examples include  sub-orders like stripes and charge density wave (CDW) \cite{Ghir,Blanko,Blackburn,Gerber,JChang}. However, it is now known that the CDW onset lies considerably below $T^*$ in underdoped samples and it is unlikely to be at the origin of the energy gap itself. Recently, one of us proposed that the pseudo-gap can be understood as a competing phase, which is a fluctuating superconductor with finite momentum, i.e. a pair density wave (PDW) \cite{Patrick}. The pairing in this phase is called Amperean pairing because it involves electron pairs on the same side of the Fermi surface, moving in the same direction. The same paper also proposed a tunneling experiment in a sandwich structure made up of optimally doped and underdoped superconductors, separated by an insulating barrier to search for evidence of this PDW phase. Given the complexity of the phase diagram, in this paper we would like to concentrate on studying the nature of the superconducting fluctuations in the pseudo-gap region.\\

\begin{figure} \hspace{-20mm}
\includegraphics[height=9cm,width=13cm]{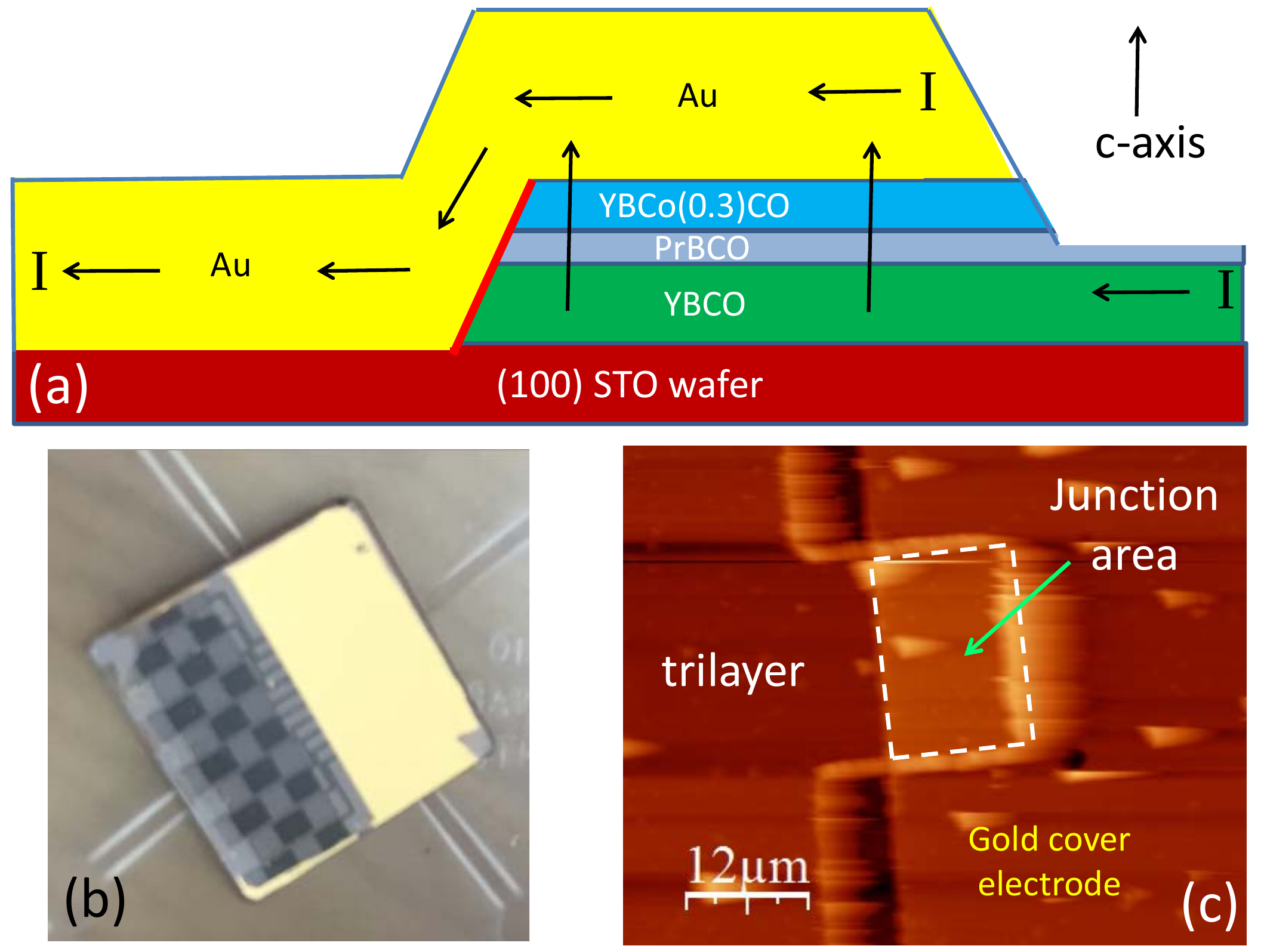}
\vspace{-0mm} \caption{\label{fig:epsart}(a) A schematic drawing of a \textit{c-axis} junction cross-section. The trilayer base electrode (of the CJ-2 wafer here) comprises of 100 nm thick $YBa_2Co_{0.3}Cu_{2.7}O_y$ on 25 nm $PrBa_2Cu_3O_{7-\delta}$ on 200 nm $YBa_2Cu_3O_{7-\delta}$, and the gold cover electrode is 400 nm thick. (b) Top view of the whole wafer. The black squares and the gold coating (yellow) are for the 4$\times$10 contacts while the 10 junctions are located in the central part of the wafer. (c) Atomic force microscope image of a single junction.
 }
\end{figure}

Motivated by these considerations, we have fabricated junctions comprised of a trilayer base electrode having a $PrBa_2Cu_3O_{7-\delta}$ (PrBCO) barrier sandwiched in between two different $YBa_2Cu_3O_{7-\delta}$ (YBCO) or doped YBCO layers, and covered by a thick gold electrode (see Fig. 1). For temperatures in between the two $T_c$'s of these junctions, early theory by Scalapino \cite{Scalapino} predicted that the \textit{c-axis} tunneling conductance will be proportional to the pair-pair correlation function $\chi$($q$, $\omega$) of the low $T_c$ electrode (thickness $d$), with $\omega=2eV/\hbar$ and $q=2\pi(\lambda_L+d/2)B/\phi_0$ where $\lambda_L$ is the London penetration length of the high $T_c$ electrode and  $\phi_0=hc/2e$ is the superconducting flux quantum. For a fluctuating PDW, the model of Ref. \cite{Patrick} predicts that one should observe a peak in the \textit{c-axis} tunneling conductance versus field at a typical field of a few Tesla which corresponds to the PDW period. It turned out that  the original model cannot be straight forwardly applied to high $T_c$ electrodes, because the formula for $q$ assumes no magnetic flux penetration into the high $T_c$ electrode. Once vortex penetration is taken into account, the magnetic field needed to observe the predicted signature of PDW pairing increases from a few Tesla to several thousand, which is beyond the depairing field, making the experiment unfeasible. This is discussed in more detail in the next section. On the other hand, while we lose the ability to obtain momentum space information, the voltage dependence of the tunneling remains a powerful tool to study any kind of fluctuating pairing, whether it is of finite momentum or not. Interestingly, we find that pair fluctuations contributing to excess current persist to temperatures much above $T_c$ in samples with a pseudogap electrode, but are practically absent in oeverdoped samples with similar $T_c$. Unexpectedly, in underdopede samples,  we find evidence of an additional channel of fluctuating superconductivity, with a lifetime much shorter than that associated with the more  conventional fluctuating superconductivity. These findings will be detailed below. We note that the conventional fluctuating peak has been reported before by Bergeal \textit{et al}. \cite{Bergeal}. Here we report of more extensive data and also contrast the observed behavior with that of junctions in which overdoped electrodes are used where no pseudo-gap is present.\\

\section{Theory}

Scalapino \cite{Scalapino} showed that pair fluctuations in the low $T_c$ electrode above its $T_c$ give rise to an excess current which is proportional to the imaginary part of the Fourier transform of the retarded pair-pair correlation function $\chi(q,\omega)$. This is derived as a linear response function, with the order parameter of the high $T_c$ electrode serving as the driving field. A magnetic field $B$ applied parallel to the junction plane supplies the momentum for the tunneling pair and gives information on the spatial correlation length, while the voltage dependence gives information on the lifetime of the fluctuating pair. The momentum supplied by the $B$ field can be seen in the following way. First, consider the case when $B$ is below $H_{c1}$ of the high $T_c$ electrode and hence is screened out by the Meissner effect. This gives rise to a screening current running along the junction. The current density  is proportional to the phase gradient according to:
 $j= 2e(n_s/m)\hbar \nabla \phi'$,
where $\nabla \phi'= \nabla \phi+i(2e/\hbar c)A$ is the gauge invariant phase gradient. We can write this as $j=(c^2 \hbar /(4\pi (2e) \lambda_L^2)) \nabla\phi'$. From Maxwell's equation, we have $j=(c/4\pi) \partial B/\partial z =(c/4\pi) B/\lambda_L$. Combining these equations we find
\begin{equation}
q= \nabla \phi' = 2\pi \lambda_L B/\phi_0.
\end{equation}
\noindent Note that compared with Scalapino's expression, we are missing the factor $d/2$. This is because Scalapino implicitly assumed phase coherence across the low $T_c$ electrode, i.e. $d$ is less than the correlation length, while we are in the opposite limit. The field penetrates fully the low $T_c$ electrode and its thickness $d$ should have no bearing on the result.\\

Once vortices penetrate the bulk electrodes, we can estimate the phase gradient in the following way. Since the cuprates are layered superconductors, the bulk vortices can be treated approximately as Josephson vortices, with core fitting between layers \cite{Koshelev}. The cross-sectional area $A'$ is equal to $A'=l_v\times l_v/\alpha$ where $l_v$ is the vortex size in the \textit{a-b} plane and $l_v/\alpha$ is its size along the \textit{c-axis}, where $\alpha$ is the anisotropy ratio ($\sim7$ for optimally doped YBCO). The area $A'$ is given by the relation $B=\phi_0/A'$. Assuming that the last layer of vortices simply terminate at the junction, we estimate the gauge invariant phase gradient to be
\begin{equation}
q'=\pi/l_v=\pi\sqrt{B/\alpha\phi_0}.
\end{equation}
\noindent Writing $q'=2\pi/L$, the length scale $L$ that can be probed by $B=10$ T is of order 800 {\AA}.  Conversely, for a pair density wave (PDW) period of $8a_0$ where $a_0$ is the in-plane lattice constant \cite{Blanko}, one has $q_{PDW}=2\pi/8a_0$. Using $q=q'=q_{PDW}$ yields the corresponding fields of $B=4.28$ T for the first case of no vortex penetration (Eq. (1)) and $B=5752$ T for the second case with vortex penetration (Eq. (2)). Since in the cuprates vortices do penetrate the superconducting electrodes for $B > B_{c1}\approx 0.01$ T, the second case is the realistic one, but its  $B$ value of 5752 T is much too high to be of any physical significance since its value is larger than the depairing field, making the experiment proposed in Ref. \cite{Patrick} unfeasible.\\

Next we review the prediction for conventional pair fluctuations. We begin with the time dependent Ginzburg-Landau free energy density
$(-i\omega/\gamma_{00} + \epsilon + \xi_0^2 q^2)|\Delta(q,\omega)|^2$ where $\epsilon=(T-T_c)/T_c$. This gives rise to
\begin{equation}
\chi = [\epsilon (-i\omega/ \gamma_0 + 1 +  \xi^2 q^2)]^{-1}
\end{equation}
\noindent where $\gamma_0 = \epsilon \gamma_{00}$ , $\xi^2 =  \xi_0^2 /\epsilon$ are the actual inverse lifetime and correlation length of the pair fluctuations. We shall take these as temperature dependent parameters from this point on.
We can re-write
$\chi(q, \omega)= \gamma_{00}[\gamma_B (1 -i\omega )/\gamma_B]^{-1}$
where $\gamma_B= \gamma_0 (1+\xi^2 q^2)$. Treating $\gamma_{00}$ as a constant, and taking the imaginary part of $\chi$, we find the current
\begin{equation}
 I(V)=A V/[\Gamma_B(1+(V/\Gamma_B)^2)]
\end{equation}
\noindent where $\Gamma_B=\hbar \gamma_B/2e$ and $A$ is an overall constant.  In a finite $B$, $q$ is given by Eq. (1) or (2) depending on vortex penetration. Thus the effect of finite B is to increase the lifetime broadening of the voltage dependence. For the realistic case of vortex penetration, we have seen that for $B=10$ T, the accessible $q$ is very small, so that we expect $q\xi << 1$ and negligible field dependence, as long as $\xi$ is less than 800 {\AA}.\\

In the following we extend the discussion to PDW fluctuations, assuming PDW at momenta $\pm \bf{Q}$. Proceeding as before except that we expand about the free energy  minima at  $\pm \bf{Q}$ , we find
$\chi = [\epsilon (-i\omega/ \gamma_0 + 1 +  \xi^2 (|\bf{q}-\bf{Q}|^2+|\bf{q}+\bf{Q}|^2))]^{-1} $ From our previous estimate, we see that for any reasonable $B$,  $|q| << |Q|$. Therefore we can write
$\chi(q, \omega)= \gamma_{00}[\gamma_Q (1 -i\omega /\gamma_Q)]^{-1}$ where $\gamma_Q=\gamma_0(1+2\xi^2|Q|^2)$. Thus we find that in the presence of a short range ordered PDW with correlation length $\xi$, there will be an excess current peaked at V=0 given by Eq. (4), except that $\Gamma_B$ is replaced by $\Gamma_Q=\hbar \gamma_Q/2e$, i.e., the width is enhanced by the factor $(1+2\xi^2|Q|^2)$. This means that a fluctuating PDW is basically indistinguishable from any other fluctuating superconductor. This is not surprising given that we do not have the momentum resolution. However, it is interesting that a PDW will make its presence felt as an excess current peak around zero voltage, as long as its coherence length is short.\\

All junctions in the present study had a $\rm T_c(high)\approx 85-90$ K and a $\rm T_c(low)\approx 50-55$ K electrodes, which allowed us to investigate pairs fluctuation currents and inverse life times in between these two temperatures. For this, we measured the conductance spectra of our junctions, and fitted the data to a pairing fluctuation model proposed by Scalapino \cite{Scalapino} and modified to include Josephson vortices in the electrodes under magnetic fields as discussed above. In addition, our junctions are described by a serial circuit consisting of the junction conductance and the serial resistance $R_0$ of the gold lead to the junction (see Fig. 1). The resulting conductance is:

\begin{equation}
G \equiv \frac{dI}{dV}=\frac{1}{R_0+G_J^{-1}(V_J,B)}
\end{equation}
\noindent where the junction conductance $G_J$ is obtained by differentiating Eq. (4) with a background conductance $G_0$ added:
\begin{equation}
G_J(V,B)=\frac{A}{\Gamma_B}\frac{1-(V/\Gamma_B)^2}{[1+(V/\Gamma_B)^2]^2}+G_0.\\
\end{equation}
\noindent The field dependent inverse lifetime $\Gamma_B$ is given by
\begin{equation}
\Gamma_B=\Gamma_0[1+\xi^2q^2(B)]=\Gamma_0(1+\xi^2\frac{\pi^2B}{\alpha\phi_0})
\end{equation}
\noindent where $\Gamma_0$ is the field independent lifetime, $\xi$ is the fluctuations correlation length, $\alpha$ is the anisotropy ratio ($\sim$7 for YBCO), $B$ is the magnetic field using $q=q'$ of Eq. (2) and $\phi_0$ is the unit flux quantum. Note that due to the non-linearity of the junction conductance,  the voltage $V_J$ across the junction enters on the right hand side of Eq. (5). It is then necessary to solve for $V_J$ as a function of $V$. To do this, we use the fact that the voltage drops on the elements of a serial circuit add up, thus
\begin{equation}
V=IR_0 + V_J.
\end{equation}
We shall see that for most of our junctions, the junction resistance is smaller than $R_0$, so that $V_J$ is 10 to 20 $\%$ of $V$. For a general nonlinear $G_J(V)$, Eq. (9) must be inverted numerically to obtain $V_J(V)$. For the special case of the Scalapino lineshape given in Eq. (6), this can be solved analytically as shown below, but it is important to note that the general features are independent of the detailed shape of $G_J(V)$. To solve for $V_J(V)$, we note that since the current $I$ through the resistor is the same as the current through the junction, we use Eq. (4) modified by the addition of a background conductance $G_0$ to re-write $V$ as
\begin{equation}
V=\frac{R_0 A(V_J/\Gamma_B)}{1+(V_J/\Gamma_B)^2}+V_J (1+G_0 R_0)\\
\end{equation}
This is an implicit equation of $V_J$ as a function of $V$. In fact, $V_J$ can be solved as the root of a cubic equation, giving $V_J(V)$.\\

Apart from the simple case when $R_0$ is small and negligible in Eq. (5), in which case the measured $dI/dV$ is simply $G_J$, another simple case is possible in the opposite limit when $R_0$ is large, provided the fluctuation conductance (first term in Eq. (6)) is small compared with $G_0$. Let us define $G_s$ = $A/\Gamma_B$ as the zero voltage value of this term. When $G_s \ll G_0$, and $R_0G_0 \gg 1$, we can expand Eq. (5) to get,
\begin{equation}
\frac{dI}{dV}=\frac{1}{(1+R_0G_0)}[\frac{A}{\Gamma'_B}\frac{1-(V/\Gamma'_B)^2}{[1+(V/\Gamma'_B)^2]^2}+G_0]
\end{equation}
where $\Gamma'_B=\Gamma_B(1+R_0G_0)$. Eq. (10) shows that in this limit, the observed conductance looks the same as the junction conductance (Eq. (6)), except that the apparent width $\Gamma'_B$ is stretched from the actual $\Gamma_B$ by a factor $(1+R_0G_0)$ which is larger than unity and the overall amplitude is decreased by the same factor. The voltage stretch factor is simply the ratio of $V/V_J$ when the junction conductance is almost linear. Most of our data in the higher temperature range is in this regime, and care must be taken to take the stretch factor into account when interpreting the inverse lifetime.
In the more general case, when $G_s \gg G_0$, the nonlinearity of the junction conductance is important. Solution of the cubic equation finds that $V_J(V)$ is nonlinear, with a small slope $1/(1+R_0G_s)$ for small voltage crossing over to a large slope $1/(1+R_0G_0)$
for large voltage. The stretch factor is now voltage dependent, giving rise to a distortion of the lineshape , eventually getting close to a top-hat shape. Our data at temperatures closer to the low $T_c$ are in this regime.

\section{Experiment.}

\begin{table}[h!]
  \centering
  \caption{\textit{c-axis} junction parameters. YBCO and PrBCO are optimally doped $YBa_2Cu_3O_{7-\delta}$ and $PrBa_2Cu_3O_{7-\delta}$, respectively and YBCoCO is underdoped $YBa_2Co_{0.3}Cu_{2.7}O_y$. All junctions were prepared on (100) $SrTiO_3$ wafers. Last column is the overlap junction area.}
  \label{tab:table1}
  \begin{tabular}{ccccc}
   \toprule[0.75pt] \toprule[0.75pt]
    wafer \# & layer 1 & layer 2 & layer 3 & area ($\mu m^2$)\\
    \midrule[0.5pt]
    CJ-1 \,\,& 300nm YBCO \,\,\,& 50nm PrBCO \,\,\,& 100nm YBCoCO\,\,\, & $ 7\times 5$\\
    CJ-2 \,\,& 200nm YBCO \,\,\,& 25nm PrBCO \,\,\,& 100nm YBCoCO\,\,\, & $20\times 15$\\
    CJ-4 \,\,& 200nm  $Y_{0.94}Ca_{0.06}Ba_2Cu_3O_y$\,\,\,& 25nm PrBCO \,\,\,& 100nm YBCoCO\,\,\, & $20\times 15$\\
    CJ-5 \,\,& 200nm YBCO \,\,\,& 25nm PrBCO \,\,\,& 100nm $Y_{0.7}Ca_{0.3}Ba_2Cu_3O_y$\,\,\, & $20\times 15$\\
    CJ-6 \,\,& 200nm $Y_{0.94}Ca_{0.06}Ba_2Cu_3O_y$ \,\,\,& 25nm PrBCO \,\,\,& 100nm $Y_{0.7}Ca_{0.3}Ba_2Cu_3O_y$\,\,\, & $7\times 5$\\
    \bottomrule[0.75pt]
  \end{tabular}
\end{table}

Fifty \textit{c-axis} junctions were prepared in the present study on five different wafers, ten junctions of the same type on each wafer, as described in detail in Table I.
The junctions structure and fabrication process are basically similar to that described previously \cite{Kirzhner}, and here we shall only give the main details. First, a whole epitaxial, cuprate trilayer was deposited \textit{in-situ } by laser ablation deposition on a $10\times 10 \times 1\,\,mm^2$ wafer of optically polished (100) $SrTiO_3$. For each CJ-i wafer, the base electrode comprised of this trilayer of layer 3 on layer 2 on layer 1 as described by each line in Table I, and shown schematically for CJ-2 in Fig. 1a. In the following step, the base electrode was patterned by Ar ion milling into ten separated bases connected to two contact pads each on half the wafer as shown in Fig. 1b. A 400 nm thick gold cover electrode was then deposited on the other half of the wafer (by a lift off process), with overlap areas on the base electrodes as given in Table I. In CJ-2, 4 and 5, no additional patterning step was done leading to a final wafer as seen in Fig. 1 (b), with large junction area as in the AFM image of Fig. 1 (c). In CJ-1 and CJ-6 however, the gold layer was coated all over the wafer, and then patterned into a cover electrode with reduced junctions area as given in Table I. It is important to note that the trilayer was deposited at high temperature of about 800 $^0$C to facilitate the epitaxial growth of the cuprate layers, while the gold layer was deposited at 150 $^0$C only. This ensured that the ion milled edges of the ten base electrodes remained damaged from the ion milling process (no reannealing at high temperature), leading to a negligible coupling of the \textit{a-b} planes to the gold cover electrode, and leaving only the good \textit{c-axis} coupling to the gold layer. Transport measurements were carried out using the four-probe technique, with or without a magnetic field of up to 8 T, parallel or perpendicular to the wafer.\\

We prepared the five sets of junctions as described in Table I, where each set had different kind of superconducting electrodes. All junctions had a low-$T_c$ electrode with $T_c\approx 50-55$ K [$\rm T_c(low)$], and a high-$T_c$  electrode with $T_c\approx 85-90$ K [$\rm T_c(high)$]. The idea was to measure mostly at temperatures in between these two transition temperatures, where fluctuations of the low-$T_c$ electrode could be probed by the order parameter of the high-$T_c$ electrode, thus enabling measurements of supercurrents and pair lifetimes in the junctions in the fluctuations regime, as discussed by Scalapino even before the cuprates were discovered \cite{Scalapino}. Moreover, except for CJ-1 and CJ-2 which have the same type of electrodes but different layers thickness, the other junctions had different kind of electrodes. CJ-1, 2 and 4 had an electrode with a pseudogap (the underdoped YBCoCO),while the counter electrode is either optimally doped (CJ-1,2) or overdoped (CJ-4).  CJ-5 and CJ-6 had no pseudogap $\rm T_c(low)$ electrode, while the $\rm T_c(high)$ side is optimally doped (CJ-5) or overdoped (CJ-6). In this way we covered all four combinations of  $\rm T_c(low)$ and $\rm T_c(high)$ electrodes, and hoped to distinguish between the different phenomena contributing to the observed results. One clear observation of the present study is that excess currents persisted above $\rm T_c(low)$ and up to $\rm T_c(high)$ only in junctions with an underdoped YBCoCO electrode which was in its pseudogap regime, while excess currents were observed only slightly above $\rm T_c(low)$ in junctions without such an electrode. This finding supports previous observation of excess currents in similar \textit{c-axis} junctions with one electrode in the pseudogap regime, as reported by Bergeal \textit{et al}. \cite{Bergeal}.

\section{Results and discussion}

\begin{figure} \hspace{-20mm}
\includegraphics[height=9cm,width=13cm]{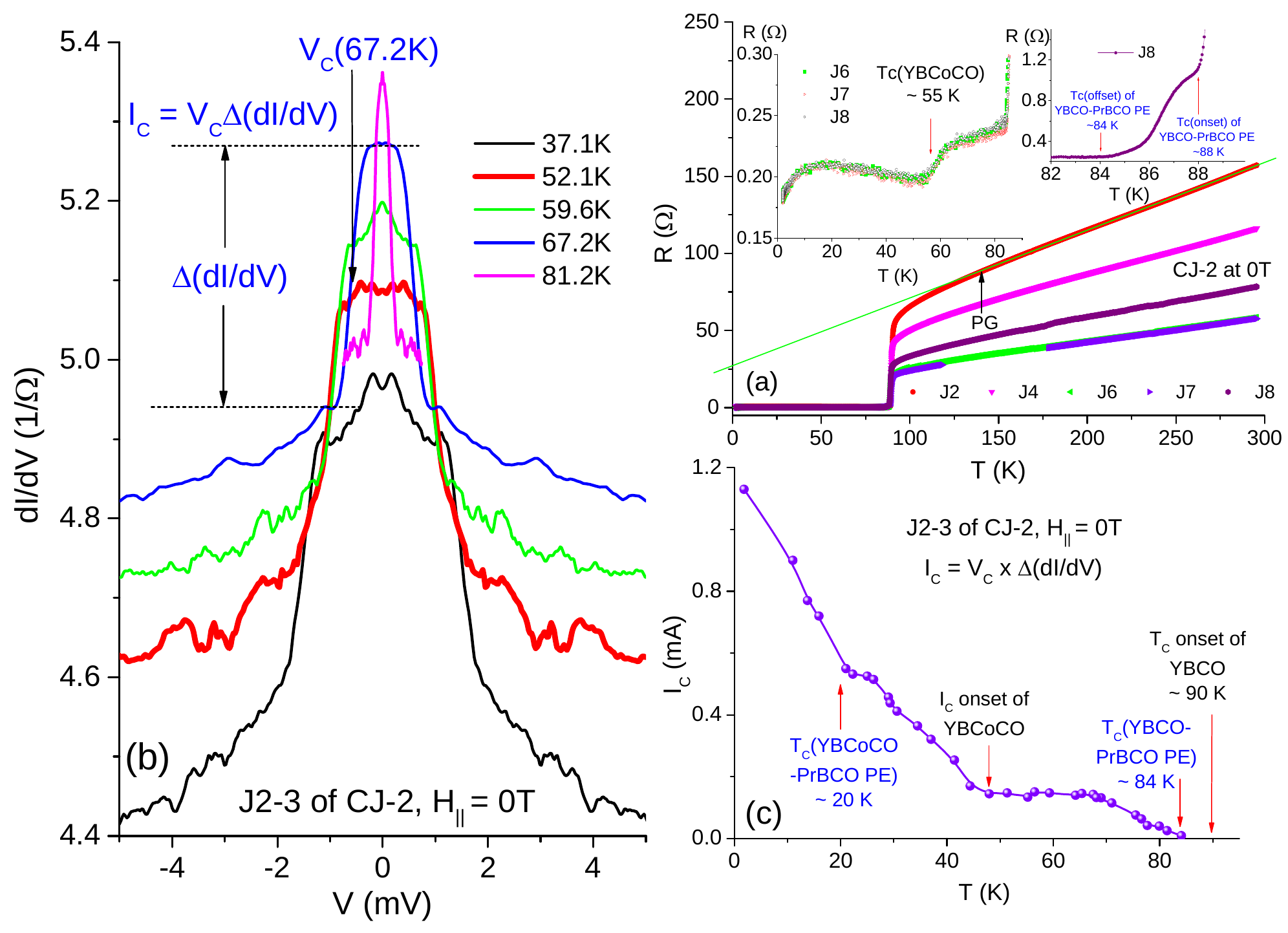}
\vspace{-0mm} \caption{\label{fig:epsart} Transport results of junctions on the CJ-2 wafer. (a) shows zero field cooled resistance versus temperature of five junctions, where the right inset shows a zoom in on the proximity region of YBCO-PrBCO just below 90 K, and the left inset a zoom in on the YBCoCO transition at 55 K. (b) shows conductance spectra of the J2-3 junctions on this wafer under zero field and different temperatures, while (c) shows the corresponding supercurrent $I_c=V_c \times \Delta (dI/dV)$ versus temperature.
 }
\end{figure}

\begin{figure} \hspace{-20mm}
\includegraphics[height=5cm,width=13cm]{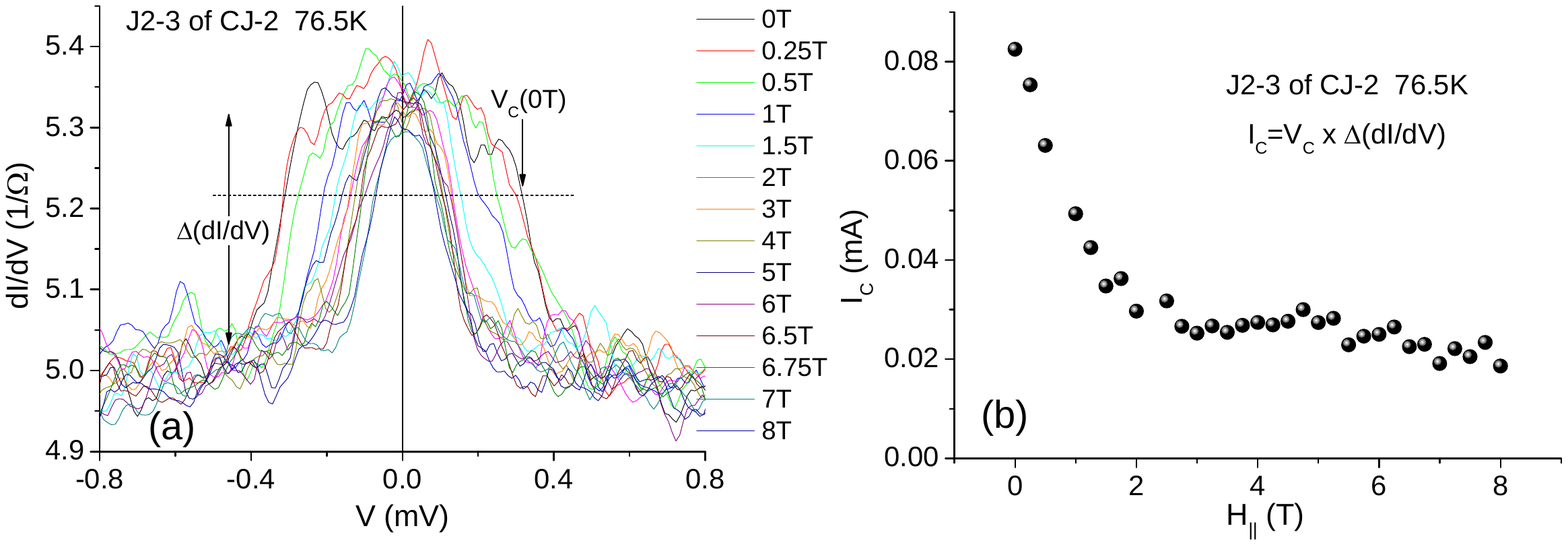}
\vspace{-0mm} \caption{\label{fig:epsart} Transport results of junctions J2-3 on the CJ-2 wafer at 76.5 K under different in-plane $H_{||}$ magnetic fields. (a) shows the conductance spectra while (b) shows the corresponding supercurrents versus field.
 }
\end{figure}

In the transport measurements of our junctions we first measured the zero field cooled resistance R versus temperature T, from which we found the transition temperatures of the different electrodes and the proximity effect (PE) regions. Then we measured the conductance spectra $dI/dV$ versus the voltage bias \textit{V} with or without magnetic fields parallel ($H_{||}$) or perpendicular ($H_\perp$) to the wafer, from which we extracted the supercurrents and inverse pair lifetimes. Since the gold cover electrode had a very small resistance, the R versus T curve above $\rm T_c(high)$ always showed the normal resistance of the YBCO or $Y_{0.94}Ca_{0.06}Ba_2Cu_3O_y$ leads to the junction in addition to the PrBCO and YBCoCO serial resistances. The result was approximately a linear R versus T as shown in Fig. 2 (a) for the CJ-2 wafer, where the deviation from linearity at about 140 K indicates the onset of the pseudogap (PG) cross-over temperature $T^*$ of the YBCoCO electrode. The right inset to this figure shows the proximity transition of the YBCO-PrBCO interface between 88 and 84 K. We can interpret this as a reduction of the effective thickness of the barrier and therefore the resistance of the junction. The left inset shows the broad YBCoCO transition at $T_c\approx$55 K, on top of the gold resistance background. Fig. 2 (b) depicts a few conductance spectra of the J2-3 junctions on this wafer under zero field and various temperatures. Most of these spectra have a top hat shape which indicates that the serial gold  resistance is dominating the junction resistance. The true junction conductance is hidden under the top hat once it exceeds the serial gold conductance. The spectra at different temperatures overlap each other because the background conductance is rising with increasing temperature. This type of spectra persisted much above $\rm T_c(low)$ and almost up to up to $\rm T_c(high)$. Since we are seeing only the "tails" of the junction conductance, its width is not an indication of the inverse fluctuation lifetime. Instead, we define $V_c$ as the voltage where the conductance $G\equiv dI/dV$ makes a sharp drop  and $\Delta(dI/dV)$ as the jump in the conductance, as indicated for the 67.2 K spectrum in Fig. 2 (a). Throughout this paper, we shall use the following operational definition of the critical current,  $I_c=V_c\times \Delta(dI/dV)$.  These critical currents were extracted from the spectra of Fig. 2 (b), and are plotted in (c) versus temperature, with the different transitions marked with arrows. Representative conductance spectra of the same junctions at 76.5 K are shown in Fig. 3 (a) under different parallel magnetic fields. The $I_c$ dependence on parallel magnetic field is depicted in Fig. 3 (b), which shows a fast decrease with increasing field up to 2 T, with a much slower decay above it. We note in passing that the above definition of $I_c$ is somewhat qualitative, as it depends strongly on $R_0$. Furthermore, the observed  voltage is not the same as the voltage across the junction due to the serial resistance as shown in Eq. (9) and corrections will be needed, as discussed later on. Therefore, one should look here only at the relative temperature and field dependencies of $I_c$ and not at the absolute values. The main result here is given in Fig. 2 (c) where  the  excess current $I_c$  is seen to persist above $\rm T_c(low)$ of YBCoCO (at about 55 K \cite{KorenPolturak}), and up to 84 K which is a few K below $\rm T_c(high)$ of YBCO (90 K). This supports the precursor superconductivity scenario in the underdoped YBCoCO electrode above its $T_c$, where fluctuating pairs tunnel through the PrBCO barrier into the YBCO electrode, leading to the observed excess current \cite{Kivelson}.\\

\begin{figure} \hspace{-20mm}
\includegraphics[height=9cm,width=13cm]{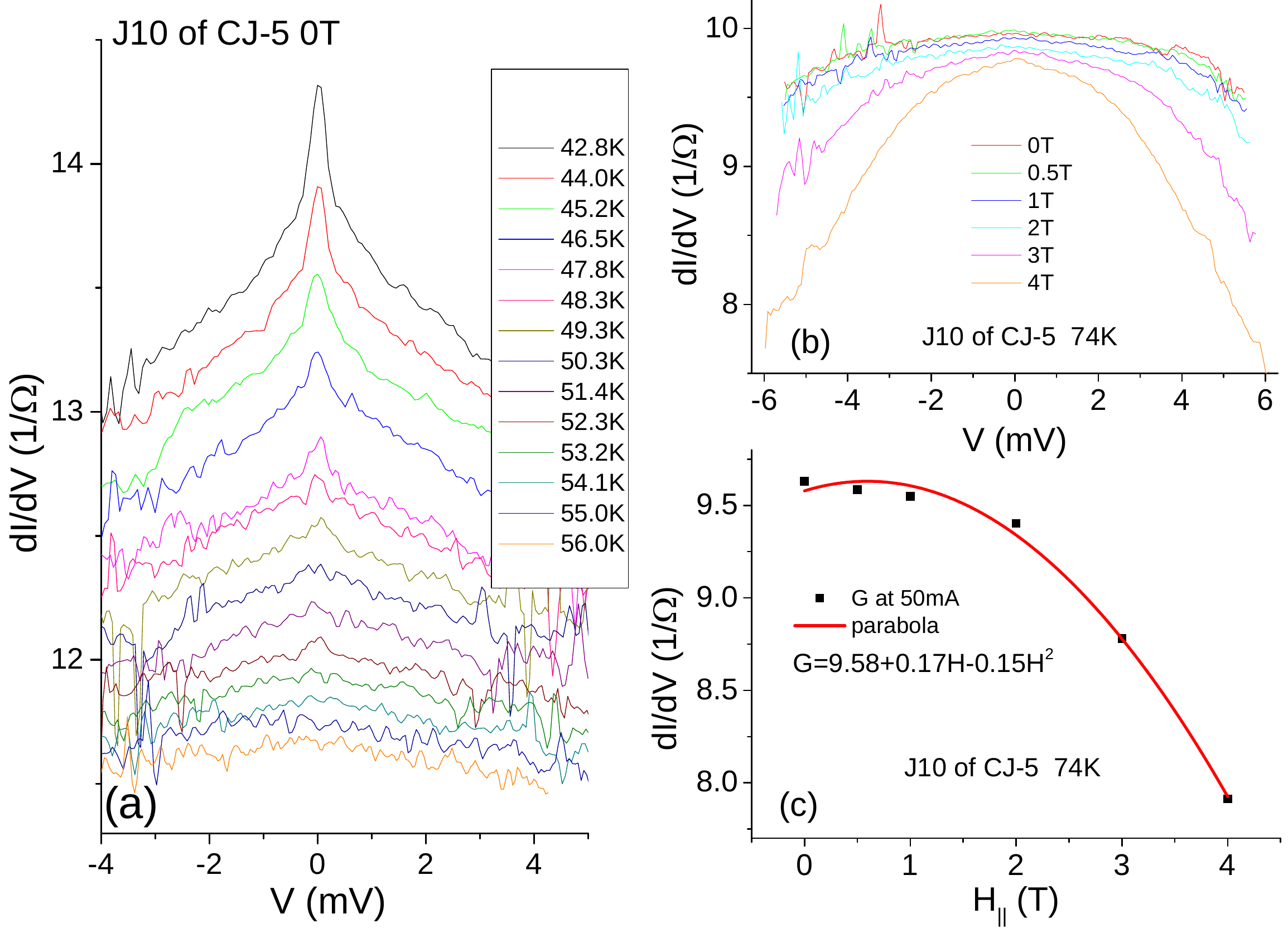}
\vspace{-0mm} \caption{\label{fig:epsart} (a) shows conductance spectra at different temperatures of the J10 junction on the CJ-5 wafer. The $T_c$ of the overdoped electrode is estimated to be 54 K from the junction resistance drop versus temperature. (b) depicts the conductance spectra at 74 K under different parallel magnetic fields. In (c) the conductance values of (b) at a constant 50 mA bias current (at about 5 mV) are plotted versus $H_{||}$, with a parabolic fit. The dominant $H^2$ term in this fit indicates the existence of flux flow conductivity originating in supercurrent in the YBCO lead to the junction. }
\end{figure}

Further support to this conclusion is found in the null results measured  above $\rm T_c(low)$ in junctions on the CJ-5 and CJ-6 wafers (see Figs. 4 and 5), where no pseudogap electrode was present (Table I), and no significant supercurrent was observed above $\rm T_c(low)$. Fig. 4 (a) depicts conductance spectra of a junction under increasing temperature. The conductance  shows a narrow peak which decreases in magnitude and appears to broaden and disappear at about 55 K, near the $T_c$ of the overdoped electrode.  This indicates that there is no excess current in this junction above these temperatures. We note that this behavior is completely different from the one observed on the CJ-2 wafer of Fig. 2, where the supercurrent persists up to 84 K (see Fig. 2 (c)). A direct comparison between the temperature dependencies of the conductance spectra of the CJ-5  and CJ-2 wafers (Fig. 4 (a) and Fig. 2 (b), respectively), shows that the narrow conductance peak of the J2-3 junctions on the CJ-2 wafer persists up to 81 K, which is about 30 K above that of the junction on the CJ-5 wafer. \\

As we go to higher temperatures, a broad peak develops as seen in Fig 4 (b). This peak becomes prominent in the presence of a parallel magnetic field. Importantly, the peak develop by a \textit{suppression} of the conductance at higher voltage. This is a well known phenomenon attributable to flux flow reduction of the supercurrent in the $\rm T_c(high)$ YBCO lead of the junction when voltage is applied. It should not be confused with an excess current $I_c$ in the junction itself. To support this claim, we measured spectra under different magnetic fields and observed  a clear decreasing conductance at high voltage bias with increasing field. Fig. 4 (c) shows conductance data taken from Fig. 4 (b) under a constant 50 mA bias current, plotted versus $H_{||}$, together with a parabolic fit. The fact that the data shows a dominant quadratic behavior is indicative of flux flow \cite{FF}. Moreover, indications that the flux flow response originates in the stronger leads rather than the junction itself, come from the behavior at even higher temperatures and higher bias currents where Larkin Ovchinikov instability and thermal runaway jumps in the I-V curves were observed \cite{LO,Beena}.\\

\begin{figure} \hspace{-20mm}
\includegraphics[height=9cm,width=13cm]{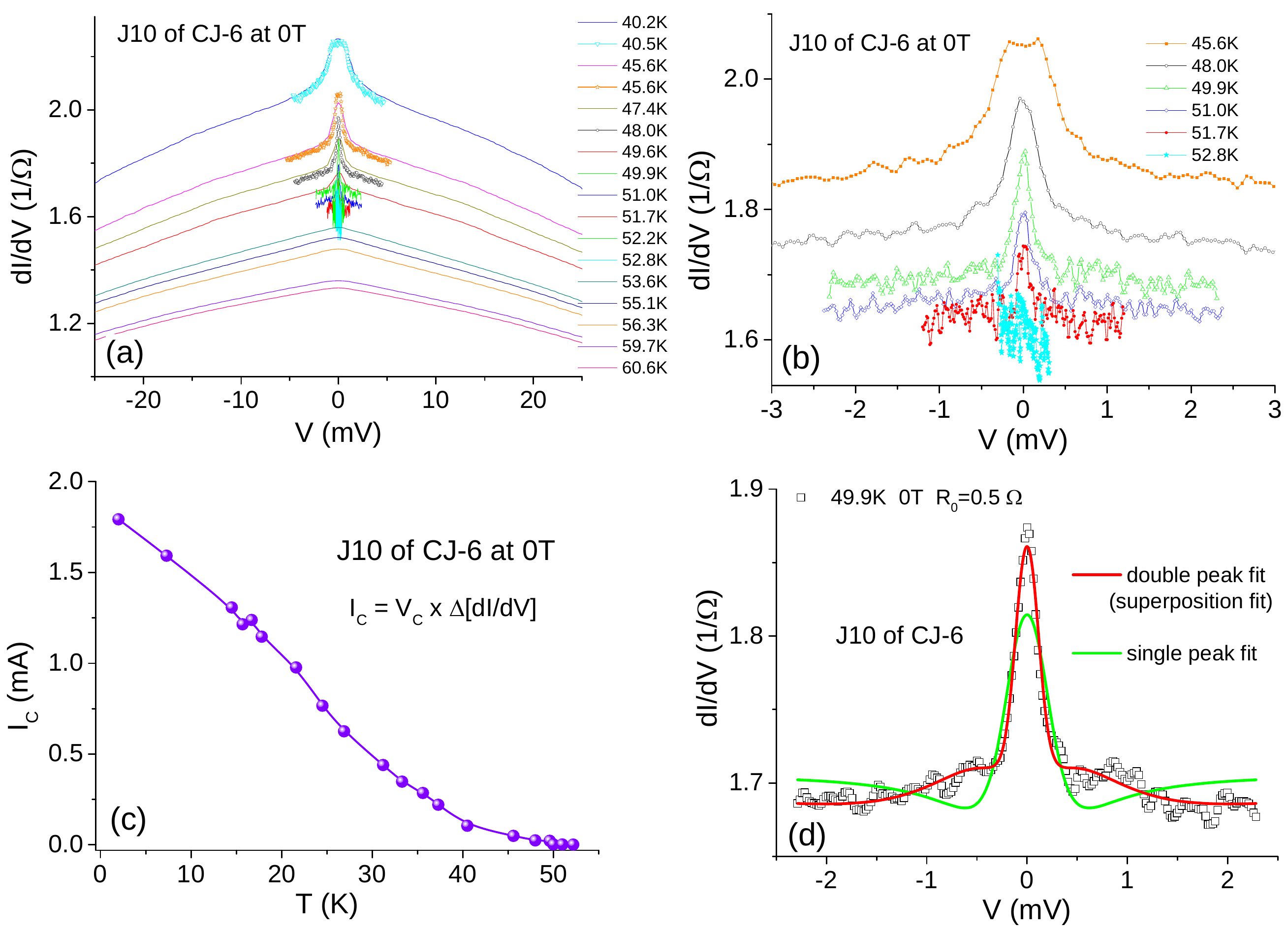}
\vspace{-0mm} \caption{\label{fig:epsart} Conductance spectra at zero field and different temperatures of junction J10 on the CJ-6 wafer are shown in (a). (b) is a zoom-in on spectra of (a) at high temperatures, where the top-hat peak at 45.6 K narrows down with increasing T and eventually vanishes at 52.8 K. (c) depicts the excess current $I_c=V_c\times\Delta(dI/dV)$ extracted from (a) and (b) as described in Fig. 2 (b). In (d), the low-V spectrum of (b) at 49.9 K is fitted using Eqs. (5) and (6) for a single fluctuations peak and a superposition of two such peaks. See Fig. S2 of the supplementary for details.  }
\end{figure}

Fig. 5 (a) and (b) depict conductance spectra of a junction on the CJ-6 wafer at different temperature and zero field. (Recall that this differs from the CJ-5 wafer only in that the $T_c(high)$ electrode is overdoped instead of optimally doped.)At 40.2 K, a clear top-hat structure is seen which narrows down with increasing temperature, until it vanishes at around 53 K. From the sharp voltage drops of these spectra the critical current $I_c=V_c\times \Delta(dI/dV)$ was extracted as shown in Fig. 2 (b), and plotted versus temperature in Fig. 5 (c). The clear difference between the supercurrent in Fig. 2 (c) and Fig. 5 (c) is that in the former where an electrode with a pseudogap is present, $I_c$ extends up to 84 K which is a few degrees below $\rm T_c(high)$, while in the latter where there is no electrode with a pseudogap it terminates at about $\rm T_c(low)\approx 50$ K. As discussed earlier, the CJ-5 wafer which also did not have a pseudo-gap electrode behaves very similarly to CJ-6.  We thus conclude that the $I_c$ above $\rm T_c(low)$ in Fig. 2 (c) is a pair fluctuations current which originates in the pseudogap regime. Nevertheless, very close to $\rm T_c(low)\approx 50$ K of the junction on the CJ-6 wafer, a pair fluctuations current still exists. This is shown in Fig. 5 (d) where a conductance spectrum at 49.9 K is presented together with two fits. One fit is to a single peak using Eqs. (5) and (6), while the other is a fit to a superposition of two peaks as in Eq. (6), but with two different amplitudes ($A_0$ and $A_1$), two different widths ($\Gamma_0$ and $\Gamma_1$) and one $G_0$. The data clearly agrees better with the second fit, which of course has the benefit of having more parameters. One reason for the better fit to a double peak may be that the line-shape predicted by Scalapino (first term in Eq. (6)) crosses zero at $V=\Gamma_B$ and has a dip beyond that voltage. Our data do not show this dip. The fit with the two amplitudes has the effect of filling in this dip which gives a better fit to the data. Thus the double peak may simply imply that fluctuations are not well described by a single Scalapino lineshape.\\

\begin{figure} \hspace{-20mm}
\includegraphics[height=5cm,width=13cm]{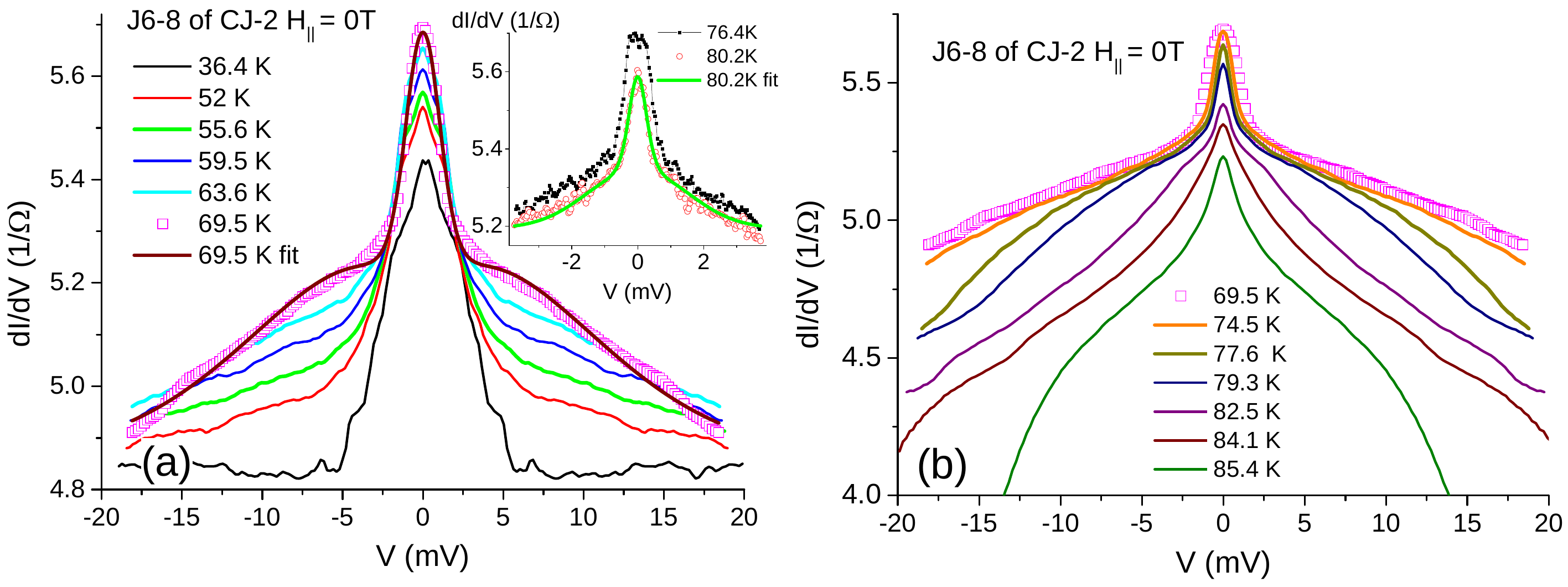}
\vspace{-0mm} \caption{\label{fig:epsart} Conductance spectra of junctions J6-8 of CJ-2 under zero field at different temperatures showing the development of the broad background peak with increasing temperature. The inset shows that under a low-V scan, the shape of the near zero bias peak is changed from top-hat at 76.4 K into a standard fluctuations peak at 80.2 K \cite{Scalapino}. Both fits at 69.5 K in (a) and at 80.2 K in the inset are fits to a double peak with the details given in the text and supplement.
}
\end{figure}

We now return to study in greater details the excess currents in junctions with a pseudo-gap electrode. So far we have focused our attention on a relatively small voltage range. By going to higher voltage we discovered that what looked like a constant background is actually the top of a broad peak of excess conductance. This is seen clearly in the temperature dependent data of CJ-2, plotted up to 20 mV in Fig. 6.  One can see that the spectrum at the lowest temperature of 36.4 K comprises of a peak on top of a flat background. With increasing temperature, this background conductance \textit{increases} and develops into a full broad peak, co-existing with the narrow central peak discussed earlier. (The rising background was noted earlier in connection with Fig. 2 (b).)  This is in contrast with samples with overdoped $\rm T_c (low)$ electrodes, where the background always decreases with increasing temperatures (see Figs. 4 (a) and 5 (a)).  On the other hand, Fig. 6 (b) shows that above $\sim$70 K, the high voltage part of the broad peak begins to drop, so that the peak appears to get narrower at higher temperatures close to $\rm T_c(high)$.  This however, is due to flux flow in the $\rm T_c(high)$ electrode, as explained earlier in connection with Fig. 4 (b) and (c). The inset to Fig. 6 (a) shows that under a low-V scan the top hat of the narrow peak of the spectrum at 76.4 K changes into a conventional peak at 80.2 K, while the shape of the background peaks remains unchanged.  We fitted the low voltage peak of the 80.2 K data with two narrow widths taking into account the fact that the voltage drop across the junction is much less than the measured voltage. The details are given in the Supplement \cite{Supplementary},  but the basic result is independent of fitting details and can be inferred by inspecting the data shown in the inset to Fig 6 (a). Recall that the conductance at large voltage is given by $(R_0+1/G_0)^{-1}$ while the top hat at zero voltage gives us roughly $R_0^{-1}$. The fact that these numbers differ by about $10\%$ in the inset to Fig. 6 (a) means that the background junction conductance $G_0$ is about 10 times $R_0^{-1}$ and hence only $10 \%$ of the voltage drop occurs across the junction. We obtained values of $\Gamma_0=0.061$ mV, $\Gamma_1=0.358$ mV which are much less than what one obtains by reading off the width in the inset. Our data support the notion that the top hat conductance peak evolving to a relatively narrow fluctuation peak is the expected pair fluctuation excess current \cite{Scalapino,Bergeal}, while the broad background peak is due to another fluctuating pair excess current of unknown origin. The existence of a broad background peak has been noted by \cite{Bergeal}, who interpreted it as being due to Andreev reflection from localized states in the barrier \cite{Davyatov}. If this were the case, this would be a property of the $\rm T_c(high)$ electrode and its interface with the insulating barrier. By comparing different combinations of electrodes, our finding that the broad peak of excess conductance is associated with the pseudo-gap electrode and not the counter electrode, effectively rules out this interpretation. Here we note that at $\sim$10 K above $\rm T_c(low)$ in Ref. \cite{Bergeal} the junction conductance is  about $0.07\,\Omega^{-1}$ while in the present study it is about two orders of magnitude higher. As the gold serial resistance $R_0$ in both studies is comparable, it turns out that in their case $V_J\approx V$, while in our case $V_J\approx V/10$. \\

\begin{figure} \hspace{-20mm}
\includegraphics[height=9cm,width=13cm]{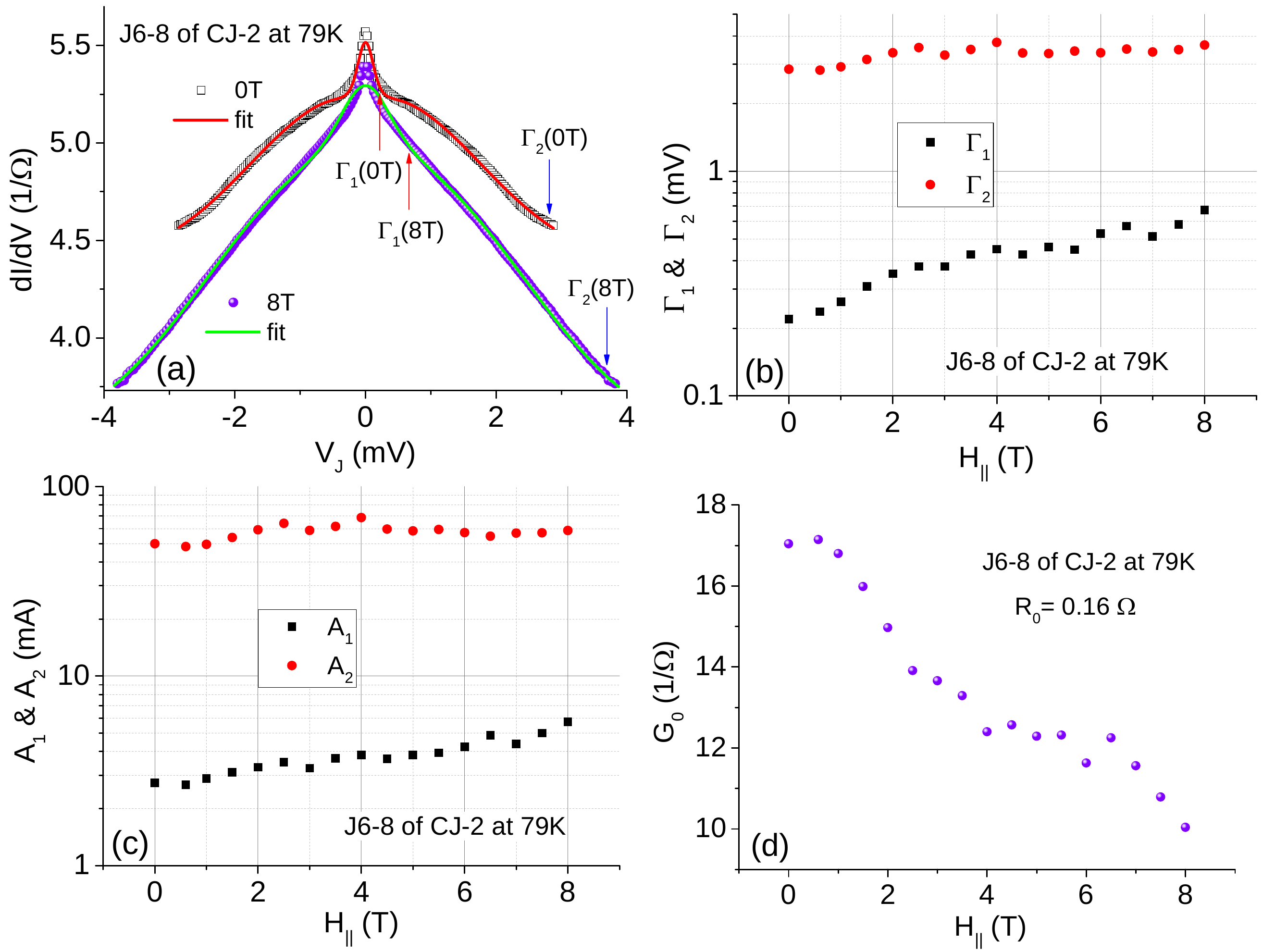}
\vspace{-0mm} \caption{\label{fig:epsart} (a) conductance spectra at 79 K of the J6-8 junctions on the CJ-2 wafer under 0 and 8 T in-plane magnetic fields versus the voltage drop on the junctions $\rm V_J$ after solving Eq. (9) (first iteration), together with the corresponding fits to a superposition of two peaks as in Eq. (6). (b)-(d) summarize the results of the fit parameters versus field (a constant $R_0=0.16\,\Omega$ was used). Raw conductance spectrum at a similar temperature of 79.3 K and 0 T can be seen in Fig. 6 (b). Clearly, there is a strong compression of the V-scale of the raw data compared to the $\rm V_J$-scale here, by a factor of about 6-7.}
\end{figure}

Next, the in-plane magnetic field dependence of our conductance spectra is presented and discussed. Fig. 7 shows data and fit analysis of results of junctions J6-8 on the CJ-2 wafer at 79 K, above $T_c$(low) in the fluctuation regime of the YBCoCO electrode (the pseudogap regime with $T_c$(YBCoCO)$\sim$55 K \cite{KorenPolturak}) and below $T_c$(high)$\sim$90 K of the YBCO electrode. We note that according to Eq. (4) the effect of a parallel field appears only as a field dependent increase of the width $\Gamma$. Furthermore, unlike the quadratic field dependent inverse lifetime of Scalapino \cite{Scalapino}, here $\Gamma_B$  of Eq. (7) is expected to be only linear in $B$. This result originates in Josephson vortices penetration into the superconducting electrodes of the junctions, a fact which was ignored in the original treatment. Moreover, for a reasonable value for the pairs fluctuation coherence length $\xi$, say of 5 nm, and the maximum field in the present study (8 T), one finds that the $B$ term in Eq. (7) is about 0.14 which is much smaller than 1. We therefore expect $\Gamma_B$ to be weakly dependent on B, except near $\rm T_c(low)$ where the coherence length $\xi$ can becomes large.  Fig. 7 (a) shows two conductance spectra of the CJ-2 wafer at 79 K under two representative fields of 0 and 8 T versus the voltage drop $V_J$ on the junction where $V_J$ is determined by the procedure described in detail in the Supplement \cite{Supplementary}.  The original raw data versus $V$ at a very similar temperature can be seen in Fig. 6 (b). As noted before, in this case $V_J$ is almost linear in $V$ and is approximately equal to 0.1 $V$. The fits to a superposition of a narrow and a broad peak are done as before in Fig. 6 (a) and its inset, except that the central peak is now fitted with a single $\Gamma_1$. The results shown in Fig. 7 are after the first iteration of solving Eq. (9) \cite{Supplementary}. The fits quality is quite good, while the only misfit occurred very close to zero bias where additional supercurrent contribution exists as in Fig. S1 \cite{Supplementary}. Fig. 7 (b-d) depict the fit parameters as a function of field. The important result here is that the inverse lifetime values separate into two distinct groups which vary slowly with field. In particular, note that  $\Gamma_2$ and $A_2$  are almost constant versus field. To summarize, one can conclude that the observed spectra indicate the existence of two inverse fluctuation lifetimes, a short one ($1/\Gamma_2$) and a long one ($1/\Gamma_1$) which have not been reported before. As explained earlier, we believe  that the broad peak is intrinsic to the pseudo-gap electrode of the junction. We see evidence of the broad peak also in CJ-4 where the counter electrode is overdoped, but due to sample degradation, we were not able to obtain reliable data up to high voltage.\\

\section{Conclusions}

Pairs fluctuation currents were investigated in $c-axis$ junctions of the cuprates with two different $T_c$ values, under in-plane magnetic fields. In junctions with a pseudogap electrode under zero field, a supercurrent was observed which persisted at temperatures above $\rm T_c(low)$ in the pseudogap regime, and up to about $\rm T_c(high)$. No such supercurrent was observed in junctions without an electrode with a pseudogap. The measured conductance spectra at temperatures above $\rm T_c(low)$ were fitted using a model of a junction with a serial resistance. We found that the data could not be fitted to a single pairs fluctuation peak, but could be fitted well to a narrow peak superposed on a very broad peak. This yielded two distinct pairs fluctuation lifetimes in the pseudogap electrode up to about $\rm T_c(high)$. Under in-plane fields, these two lifetimes remained separated in two distinct groups. The long lifetime varied with increasing field moderately while the short lifetime is almost field independent which may indicate that a short coherence length is associated with this pairs fluctuation.\\

The observation of two kinds of pair fluctuations above $T_c$ reminds us of the conflicting reports in the literature that some measurements indicate that pair fluctuations is limited to 20 K above $T_c$ \cite{Orenstein}, while others find evidence for it up to 180 K \cite{Keimer}. It is tempting to identify the broad excess current peak which is independent of temperature and parallel magnetic field with the fluctuations observed in the latter experiment. Of course, this leaves open the microscopic origin of this pairs fluctuation. Here we simply mention two candidates in the literature, and there are certainly others. First, Geshkenbein, Ioffe and Larkin \cite{Geshkenbein} have proposed a picture of preformed pairs made up of electrons near the anti-nodes (0,$\pi$) which have a small dispersion. In their picture these preformed pairs co-exist with more conventional BCS pairing made up of nodal electrons. Thus, two kinds of pairs leading to two different lifetimes. Second, there is the proposal by one of us \cite{Patrick} that a fluctuating pair density wave PDW is responsible for the pseudo-gap, and as discussed in section 2, this may show up as a conductance peak as a function of voltage, even though the pair momentum is too large to be measurable by our available magnetic field. This conductance peak would a-priori have different lifetime than that of conventional fluctuations, thus in the experiment two distinct pairs fluctuation lifetimes will be observed as actually found in the present study. In either scenarios, it is possible that the second kind of pair fluctuation also becomes coherent below $T_c$, which would explain why this fluctuation seems to emerge above $T_c$, taking weight from a narrow central peak of supercurrent which is obscured by the top hat conductance.  While our finding does not distinguish between different models, we believe the very existence of two types of pair fluctuations with very different lifetimes is a significant development.\\

{\em Acknowledgments:} G. K. thanks Amit Keren for pointing out Ref. \cite{Patrick} which started this whole project. PAL acknowledges support by NSF under DRM-1522575.

\bibliography{AndDepBib.bib}

\bibliography{apssamp}

\newpage

\begin{center}
{\large  \textbf{Supplementary material}\\
\vspace{5mm}
for\\
\vspace{5mm}
\textbf{Observation of two distinct pairs fluctuation lifetimes and\\
 supercurrents in the pseudogap regime of cuprate junctions}\\
 \vspace{5mm}
 by\\
 \vspace{5mm}
 \textbf{Gad Koren and Patrick A. Lee}\\
 }
\end{center}

\normalsize \baselineskip=8mm

\newpage


\textbf{In this supplementary part we describe in details the fitting procedure to obtain the voltage drop across the junction $V_J$ as a function of the measured voltage $V$, in a  model of a junction with a serial resistor $R_0$ obeying Eq. (9) of the main article.}\\

As explained in the theory section, due to the fact that our junctions have a serial resistance ($R_0$), the actual voltage drop on the junction $V_J$ is smaller than the measured $V$ as is obvious from Eq. (8). Therefore, we cannot read off the width of the spectrum directly from Fig. 5 (d) to extract an inverse lifetime. We need to determine $V_J$ as a function of $V$ first, and plot the data vs $V_J$. To illustrate this, we take a conductance spectrum of the J2 junction on the CJ-4 wafer under a low-V scan at 81.2 K and 0 T, which is well in the fluctuations regime between $\rm T_c(low)$ and $\rm T_c(high)$ (see Fig. S1 (a)). Due to a small drift in temperature during the measurement, the conductance spectrum was slightly asymmetric, and therefore symmetrized with respect to $\pm$V values, to allow for fit and comparison to simulation with symmetric functions. The spectrum has a "top hat" shape with a rather rapid drop at about 1.2 mV and a very narrow peak above the top hat. We interpret the narrow peak to be due to inhomogeneity in the junction area, which supports a small amount of supercurrent and we do not attempt to fit it. As a first approximation, we fitted this data using Eqs. (5) and (6) as  done before in Fig. 5 (d) for a single peak. The resulting fit fails to  reproduce the corners of the top-hat shape of the spectrum, but overall it is close enough to the measured data and yields a set of initial fitting parameters which are used to solve Eq. (9). This yields $V_J$ as a function of $V$, which is then used to replot the measured conductance spectrum versus $V_J$, and fit it again using Eqs. (5) and (6). This first iteration fit produced a new set of parameters which were used to solve Eq. (9) again, and the iteration process was repeated. The resulting $V_J(V)$ after the first iteration is shown in the inset to Fig. S1 (a). The initial slope is about 1/3 and becomes nonolinear beyond $V\approx 3$ mV.  Fig. S1 (a) shows a conductance spectrum simulation using Eqs. (5) and (6) with the parameters of the first iteration (red curve). The result shows a significantly narrower peak than seen in the measured data, and it shows the dips in the Scalapino line-shape in accordance with Eq. (6) \cite{Scalapino}. To compare this simulated spectrum with the raw data, we stretched its V-scale by a constant factor to have the same width as that of the original raw data, and found that it actually overlaps the whole curve of the fit to the raw data (green circles and blue curve). This indicates a linear scaling of the measured data with the simulated (true) spectrum for V less than 3 mV, consistent with our earlier observation that $V_J(V)$ is linear in this range. \\

\begin{figure} \hspace{-20mm}
\includegraphics[height=6cm,width=15cm]{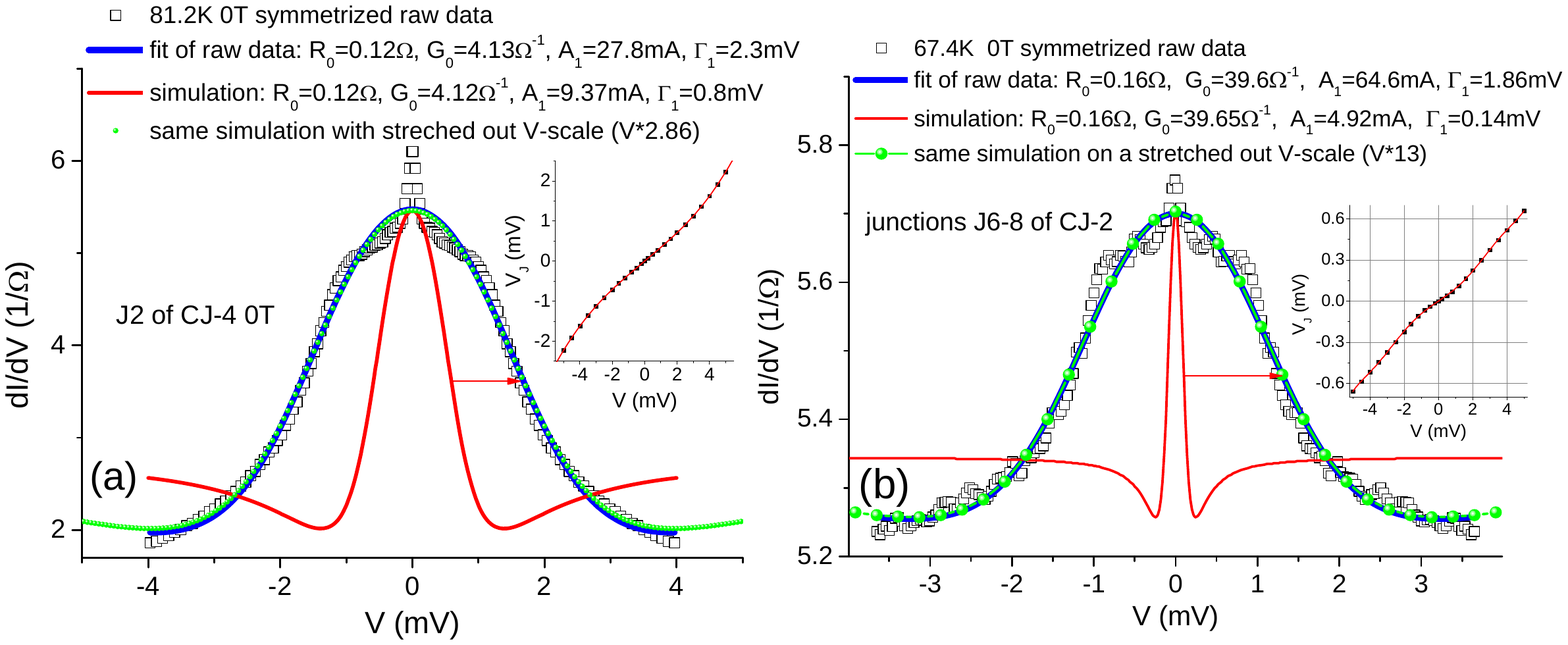}\\
Fig. S1: Conductance spectra at 0 T of junction J2 of the CJ-4 wafer at 81.2 K (a), and junctions J6-8 of the CJ-2 wafer at 67.4 K (b),  together with one fit and one simulation each. The fits to the raw data are done using Eqs. (5) and (6) for a single peak. Then in an iteration process, the parameters of these fits are used to solve Eq. (9), get $V_J(V)$ (see insets including fits to polynomials of order 7), replot the spectra versus $V_J$, fit these spectra as before, and get two new sets of parameters for (a) and (b). The simulations use these first iteration fit parameters to calculate and re-plot the spectra versus the measured voltage $V$ (red curves). Then the V-scale of the simulations are stretched out by constant factors of 2.86 for (a) and 13 for (b), to coincide with the widths at half maximum of the fits to the raw data (see the arrows). As can be seen, this results in a full overlap of the whole stretched simulation curves with the first fits to the raw data in both cases.
\end{figure}

We tested the iterations procedure up to third iteration, and found convergence already after the first iteration. We thus show results of the first iteration throughout this study. Fig. S1 (b) shows results of applying the same fitting and simulation procedure to junctions J6-8 of the CJ-2 wafer at 76.4 K and 0 T. One finds that this procedure works well also for these junctions, but now with a much larger stretch factor.  Concerning the number of free parameters in our fits using Eqs. (5) and (6), we note that $R_0$ and $1/G_0$ play a similar role and are therefore dependent. Since $R_0$ in the present study is the resistance of the gold lead to the junctions, we used a constant value for it in the fits. Note that in highly transparent junctions where $R_0\gg 1/G_0$, the $R_0$ value was almost equal to the measured value obtained from the resistance versus temperature result. We therefore have only three free parameters in our fits of a spectrum with a single peak ($\Gamma$, $A$ and $G_0$).\\

\begin{figure} \hspace{-20mm}
\includegraphics[height=8cm,width=11cm]{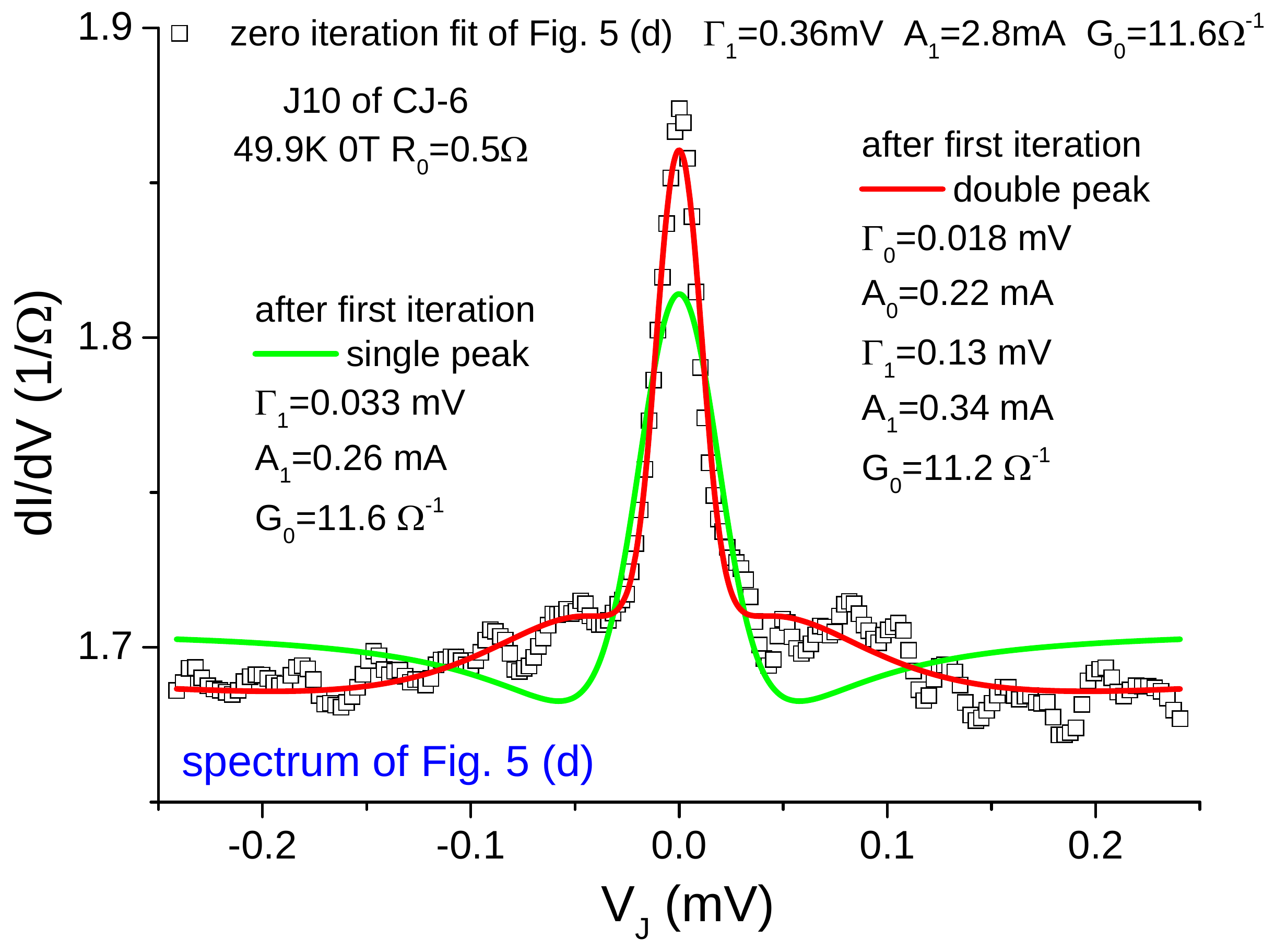}\\
Fig. S2: Raw data conductance spectrum of Fig. 5 (d) of junction J10 of the CJ-6 wafer at 49.9 K and 0 T is plotted here versus $V_J$ together with two fits. The zero iteration fit parameters are those of a single peak fit to the original spectrum of Fig. 5 (d) versus V. The two fit curves were obtained after the first iteration (solving Eq. (9)) for a single and double peaks as in Eq. (6).
\end{figure}

Returning to Fig. 5 (d), this procedure should lead to a narrower conductance spectrum than the one shown there, when the spectrum is replotted vs $V_J$.  In Fig. S2 we replotted the measured $dI/dV$ spectrum versus $V_J$ and fitted it to a superposition of two peaks as done before with the original $dI/dV$ versus $V$ data. We found $\Gamma_0=0.018$ mV and $\Gamma_1=0.13$ mV both of which are narrower by about an order of magnitude compared to the values obtained by a direct fit of the data as in Fig. 5 (d) ($\Gamma_0=0.19$ mV and $\Gamma_1=1.3$ mV). For the poorer single peak fit in Fig. S2, the numbers are $\Gamma_1=0.36$ mV for the zero iteration, and $\Gamma_1=0.033$ mV after the first iteration. Evidently for this sample, 90$\%$ of the voltage drop occurred on the serial resistor. Here we just point out that whether we fit with one or two peaks, both peak widths $\Gamma_0=0.018$ mV and $\Gamma_1=0.13$ mV are sufficiently narrow to indicate that this feature originates in a fluctuating supercurrent. In this sample we attribute it to conventional Gaussian pair fluctuations observable only slightly above $T_c$.\\

\end{document}